\documentclass{PoS}

\usepackage{gensymb}

\title{ATLAS Pixel Detector: Operational Experience and Run-1 to Run-2 Transition}

\ShortTitle{ATLAS Pixel Detector: Operational Experience and Run-1 to Run-2 Transition}

\author{\speaker{Alessandro La Rosa}, on behalf of the ATLAS Collaboration\\
        DPNC, University of Geneva\\
        CH 1211 Geneva 4, Switzerland\\
        E-mail: \email{alessandro.larosa@cern.ch}}

\abstract{The Pixel Detector of the ATLAS experiment has shown excellent performance during the whole Run-1 of LHC. Taking advantage of the long shutdown, the detector was extracted from the experiment and brought to surface, to equip it with new service quarter panels, to repair modules and to ease installation of a new innermost layer, the Insertable B-Layer (IBL).
An overview of the operational experience, the refurbishing of the Pixel Detector and of the IBL project as well as the experience in its construction, integration and commissioning are described.
}

\FullConference{The 23rd International Workshop on Vertex Detectors,\\ 15-19 September 2014\\ Macha Lake, The Czech Republic}

\begin{document}

\section{Pixel Detector}
ATLAS  \cite{1} is a general purpose experiment operating at the Large Hadron Collider (LHC) at CERN. The ATLAS detector was designed to be sensitive to a wide range of physics signatures to fully exploit the physics potential of the LHC collider at a nominal luminosity of $10^{34}$\,cm$^{-2}$s$^{-1}$. As most of the final states of collisions in the ATLAS experiment include charged particles, an excellent tracking system is essential.
\begin{figure}[h!]
\centering
\includegraphics[scale=0.6]{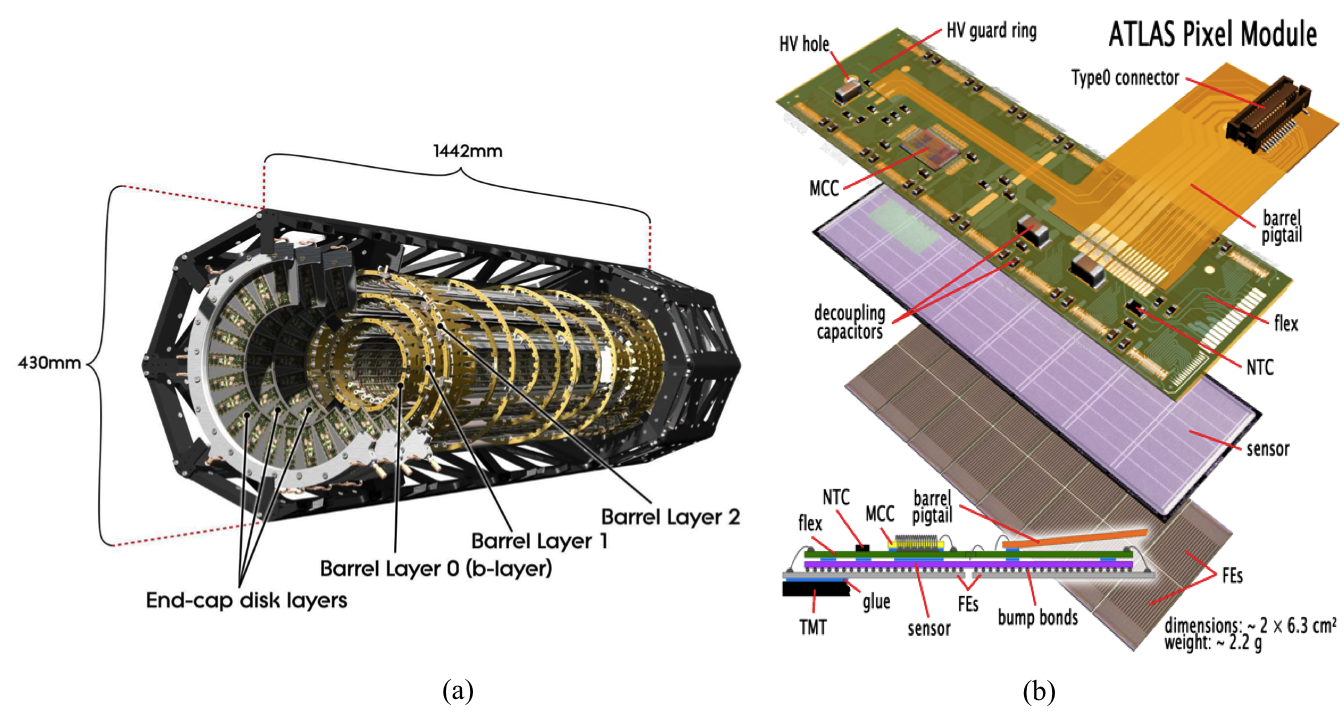}
\caption{(a) Schematic drawing of the ATLAS Pixel Detector. The detector comprises three concentric barrel layers and two end-caps with three discs each. The single detector modules are mounted on carbon fibre support structures with incorporated cooling circuits. (b) Assembly view and cross-section of an ATLAS Pixel Detector module. Sixteen front-end chips are bump bonded to the silicon pixel sensor. Interconnections are done on a flexible kapton PCB, which is connected by wire bonds to the electronics chips \cite{2}.}
\label{fig:Pixel_run1}
\end{figure} 
\par The Pixel Detector \cite{2} is the core of the ATLAS Inner Detector (ID), and as shown in Figure\,\ref{fig:Pixel_run1}, it consists of three concentric barrel layers with mean radii of 50.5\,mm, 88.5\,mm and 122.5\,mm centred around the beam axis and two end-caps with three discs each, forming a three-hit system up to pseudo-rapidities of $\pm$\,2.5. The full detector contains 1744 pixel modules, which are mounted on carbon fibre local supports (staves). An evaporative C$_{3}$F$_{8}$ cooling system is incorporated into the staves to absorb the heat produced by the modules and to allow for an operation at temperatures below 0\,$\degree$C to limit the effects of radiation damage. The individual pixel modules are made of a 250\,$\mu$m thick n$^{+}$-on-n silicon sensor, 16 front-end chips (FE-I3 \cite{3}) and a module controller chip (MCC \cite{4}). The sensor is divided into 47232 pixels with a typical pixel size of (50 x 400)\,$\mu$m$^{2}$ and approximately 10\% of pixels (long pixels) have a size of (50 x 600)\,$\mu$m$^{2}$ to bridge the gaps between the front-end chips. In addition, in each column eight pairs of pixel implants, located near the center lines, are ganged to a common read-out, resulting in 320 independent read-out rows or 46080 pixel read-out channels. The front-end chip provides pulse height measurement by means of time over threshold (ToT) with zero suppression on chip while the module event is built by the MCC. 
\newpage
\section{Operational experiences}
The successful operation of the Pixel Detector during the LHC Run-1 was possible thanks to regular tuning and calibration of the detector.
\begin{figure}[h!]
\centering
\includegraphics[scale=0.6]{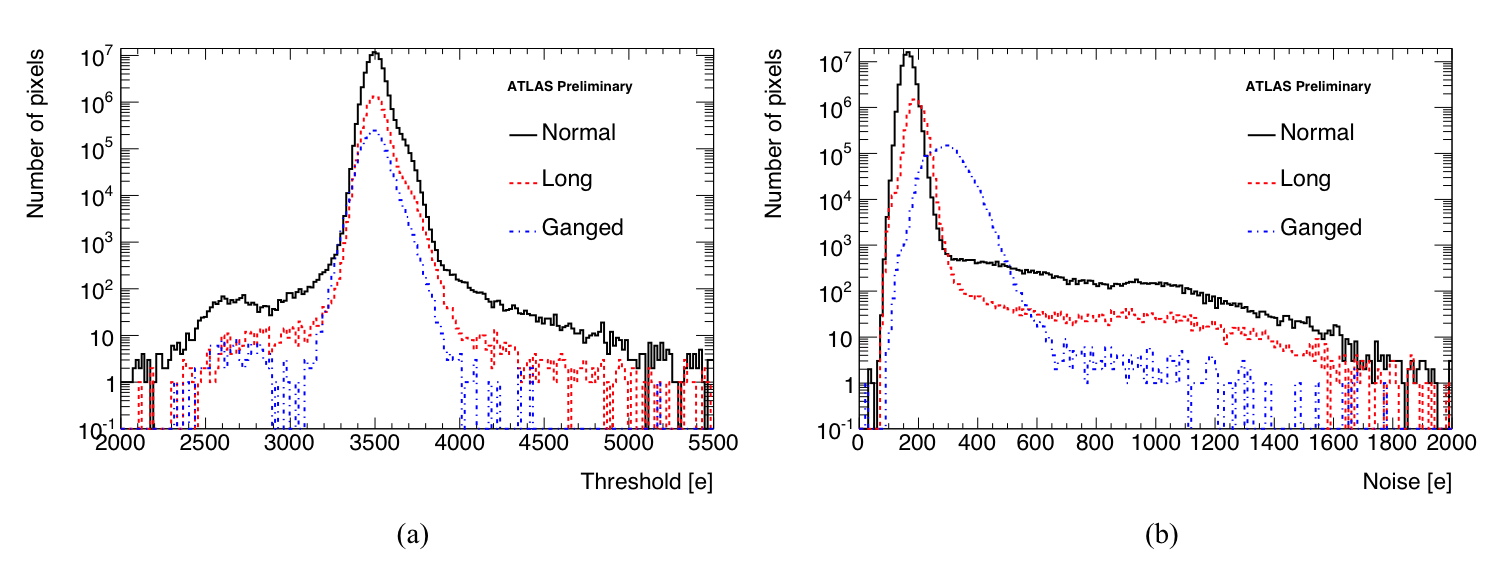}
\caption{Measured threshold values for all pixels in the detector. Prior to the scan the threshold setting have been tuned to 3500\,e$^{-}$. (b) Noise values obtained from the same threshold scan \cite{5}.}
\label{fig:TH_N}
\end{figure} 
\par The pixel thresholds were tuned to 3500\,e$^{-}$ with a typical dispersion of 40\,e$^{-}$. The measured noise was 180\,e for normal sized pixels, with a higher value for long pixels and about 300\,e for ganged pixels as expected due to the higher capacitance these pixels present to the readout electronics. Threshold and noise values obtained from a threshold scan are shown in Figure\,\ref{fig:TH_N} \cite{5}. Pixels exceeding noise occupancy of 10$^{-6}$ hits per bunch crossing in dedicated noise data taking runs were masked for data taking already in the module configuration. Typical occupancies for physics data have been in the order of 10$^{-4}$. Only 0.1\% of the pixels needed to be masked out to achieve such performance. The front-end chip has the capability to provide information about the deposited charge by the measured Time Over Threshold (ToT), which has a nearly linear dependence on the charge released in the sensor. The tuning target was a ToT of 30 bunch crossings for an injected charge of 20\,ke$^{-}$ that corresponds to a minimum ionization particle deposited charge.
\begin{figure}[h!]
\centering
\includegraphics[scale=0.6]{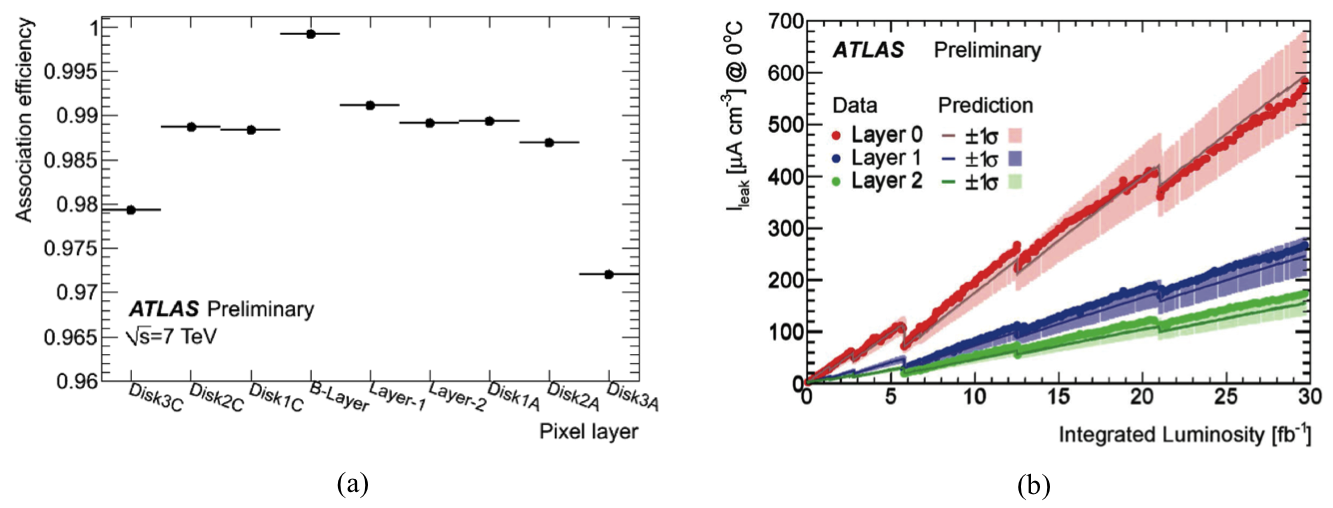}
\caption{(a) Hit association efficiency for the different parts of the ATLAS Pixel Detector. The high efficiency of the B-layer is an artefact of the track set used for the measurement, whereas the low efficiency of the outermost discs is due to few bad modules known from production [5]. (b) The averaged reverse- bias current for all Pixel modules in the different barrel layers as a function of the integrated luminosity \cite{5}.}
\label{fig:Hit_Ikeak}
\end{figure} 
\par Looking at the detector performance the efficiency of tracks having associated hits in the different Pixel Detector layers was about 99\%. As shown in Figure\,\ref{fig:Hit_Ikeak}(a), a slightly lower efficiency in the outermost disks (Disk 3C of  $\thicksim$\,0.98 and Disk 3A of $\thicksim$\,0.97) was measured and it was due to few bad modules that were already identified at the production stage.
\par The radiation damage effects were within expectation and they become visible when a significant increase in leakage currents has been observed in Pixel Detector. The leakage current was mainly measured by  high voltage power supplies on a level of 6 or 7 modules per channel and with a precision of 80\,nA. As shown in Figure\,\ref{fig:Hit_Ikeak}(b), the evolution of leakage current with the respect to the integrated luminosity was in agreement with the prediction. The steps in the leakage current evolution correspond to annealing during detector warm-up periods.
\par Despite the Pixel Detector showed very good performance with high efficiency and data quality, there were specific challenges that have been faced during Run-1. At the beginning of Run-1, 97\% of the pixel modules was included in data taking, but the number of disabled or problematic modules and FEs has been continually increasing up to 88 modules and 60 FEs. Failures were highly correlated to thermal cycling. To mitigate this issue, the cooling system was continually operated whenever it was possible. In addition to that, other two main issues have been faced. One was failures in the operation of the VCSELs of the off-detector optical transmission that has been solved by replacing the failing components. The other challenge was the high module occupancy and trigger rates that leaded to the modules de-synchronization during the 2012 data taking. As mentioned the de-synchronization was strongly correlated to the occupancy and happens more often at the beginning of runs. It was probably caused by single event upsets in the digital logic of modules or by high bandwidth data bursts. To synchronize modules automatic recovery actions were implemented.

\section{Detector upgrade during the LHC long shutdown 2013-14}
Due to the location of the Pixel Detector and considering the expected lifetime of the Layer-0 (or B-Layer, which is the closest to the beam pipe) an upgrade of the detector was needed to guarantee an excellent overall performance of the ATLAS tracking system over the full lifetime up to the High-Luminosity LHC phase ($\thicksim$\,2023).
\par Taking advantage of the LHC long shutdown in 2013 and 2014 (LS1), the Pixel Detector was extracted from the experiment and brought to surface to equip it with new service quarter panels, to repair modules and to ease installation of a new innermost layer, the Insertable B-Layer (IBL) \cite{6}. The IBL is crucial to guarantee an excellent vertex detector performance and to provide redundancy in case of failures or radiation damage in the three layers Pixel Detector. In addition it was taken the opportunity to upgrade the Layer-2 DAQ hardware, by doubling the readout speed, for being ready for LHC Run-2 luminosity and beyond and install the Diamond Beam Monitor (DMB) that aims to measure the bunch-by-bunch beam luminosity.

\section{Refurbishment and consolidation of the Pixel Detector}
The main motivation to replace the service quarter panel (SQP) of the Pixel Detector was driven by the fact that during the operation of the detector, some of the lasers in the off-detector systems started to fail. The same type of laser was also used in the on-detector system (Optoboards) integrated on the SQP that cannot be accessed without removing the detector from ATLAS. Due to the Optobards inaccessibility their possible failures were a major concern for the operation of the Pixel Detector throughout its lifetime. So that it was decided to build new service panels (nSQP), which enable the Optoboards to be moved into a serviceable area by extending the original 1-meter electrical readout by an additional 6.6 meters. The nSQP project allowed also to increase the bandwidth of Layer-1 to be the same as Layer-0 (160\,Mbit/s). This increase in readout speed allows to move more data out of the detector, making it suitable for operation at higher luminosities as currently predicted in the LHC schedule.
\par In April 2013 the Pixel Detector was extracted from the experiment and brought to surface where the SQPs were removed from the detector, and an intensive and detailed investigation of the failing modules was carried out.  In those cases where it was possible the module disconnections were repaired. Once the nSQPs were installed the Pixel Detector was reassembled and tested. In December 2013 the detector was reinstalled in the ATLAS experiment. In May 2014 all the services were connected and after having retested the detector the module recovery went from 95\% (at the end of Run-1) to 98\% of working modules. Most of the significant improvements took place in the Layer-0 where the failing modules were reduced from 6.3\% to 1.4\% and in the Layer-2 where the failing modules were reduced from 7\% to 1.9\%. 

\section{The Insertable B-Layer (IBL): from the deign to the commissioning}
The IBL Detector is the additional fourth pixel layer that has been built around the new Beryllium beam pipe (4\,mm of radius) and then inserted inside the Pixel Detector in the core of ATLAS detector.  It consists of 14 carbon fibre staves 64\,cm long, 2\,cm wide, and tilted in $\phi$ of 14\degree  surrounding the beam-pipe at a mean radius of 33\,mm and a pseudo-rapidity coverage of $\pm$\,3. Each stave, with CO$_{2}$ cooling integrated, is equipped with 32 front-end chips (FE-I4 \cite{6}) that are bump bonded with the Silicon sensors. 
\begin{figure}[h!]
\centering
\includegraphics[scale=0.15]{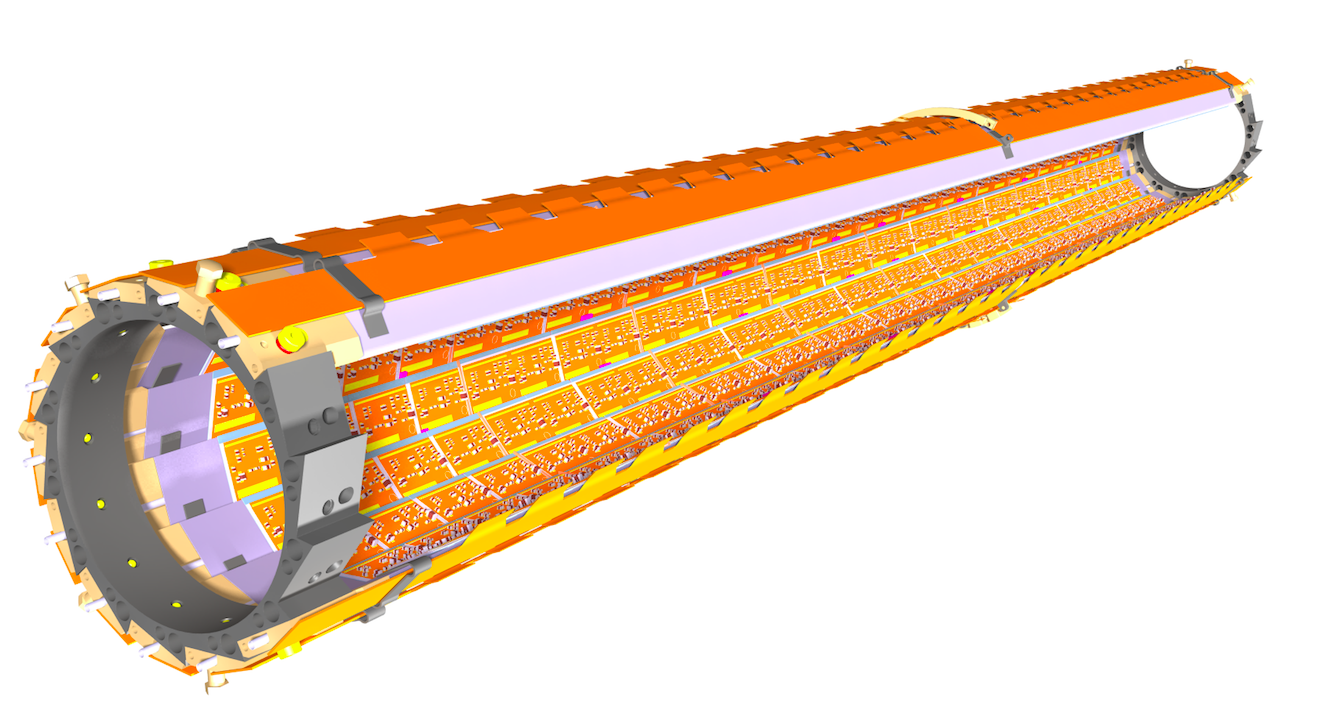}
\caption{Schematic drawing of the ATLAS IBL Detector. The single detector modules are mounted on carbon fibre support structures with incorporated CO$_{2}$ cooling circuits.}
\label{fig:IBL}
\end{figure} 
\par The FE-I4 chip is designed in 130\,nm CMOS technology and it consists of 26880 pixel cells organized in a matrix of 80 columns (on 50\,$\mu$m pitch) by 336 rows (on 250\,$\mu$m pitch). Each front-end cell contains an independent, free running amplification stage with adjustable shaping, followed by a discriminator with independently adjustable threshold. The FE-I4 keeps tracks of the firing time of each discriminator as the time over threshold with 4-bit resolution, in counts of an external supplied clock of 40\,MHz nominal. A common sensor baseline for engineering and system purpose was chosen for the pixel module considering that there are two different silicon sensor technologies: planar n$^{+}$-on-n \cite{6} manufactured by CiS (Germany) and 3D with passing through columns \cite{6} manufactured by FBK (Italy) and CNM (Spain). The basic unit of the IBL is a module that consists of two (for planar) or one (for 3D) front-end chips bump bonded to one sensor tail. For single-chip (two-chip) assemblies the nominal active coverage for particles normal to the beam is 98.8\% (97.4\%). Looking at the stave layout twelve two-chip planar modules cover the central part of the stave while four single-chip 3D modules cover the forward regions of the both ends of the stave. The forward regions of the staves are populated by 3D modules where they can guarantee a better z-resolution in the tracking reconstruction after heavy irradiation due to their electrodes orientation. Considering the high radiation levels expected for the IBL (NIEL dose: 5\,$\times$\,10$^{15}$ 1\,MeV n$_{eq}$\,cm$^{-2}$ and TID: 250 Mrad) all the detector components have been properly qualified up to such fluence.
\par
\begin{figure}[h!]
\centering
\includegraphics[scale=0.5]{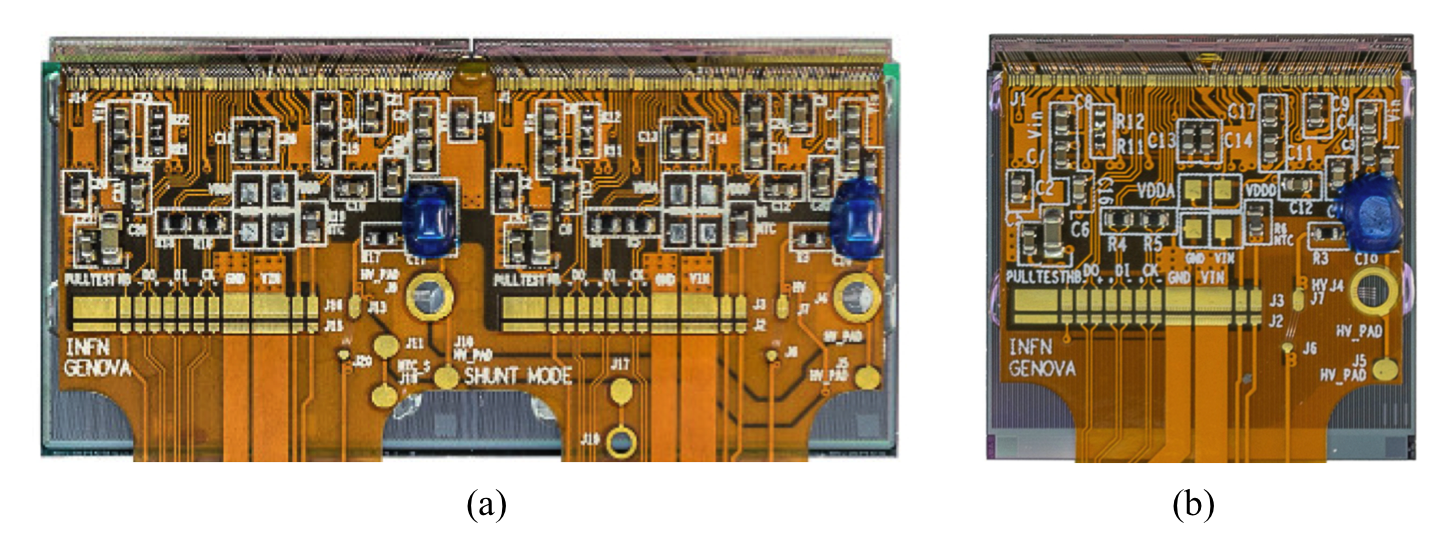}
\caption{Photo of a dressed IBL two-chip (a) and a single-chip (b) module.}
\label{fig:IBL-modules}
\end{figure}

\begin{figure}[h!]
\centering
\includegraphics[scale=0.3]{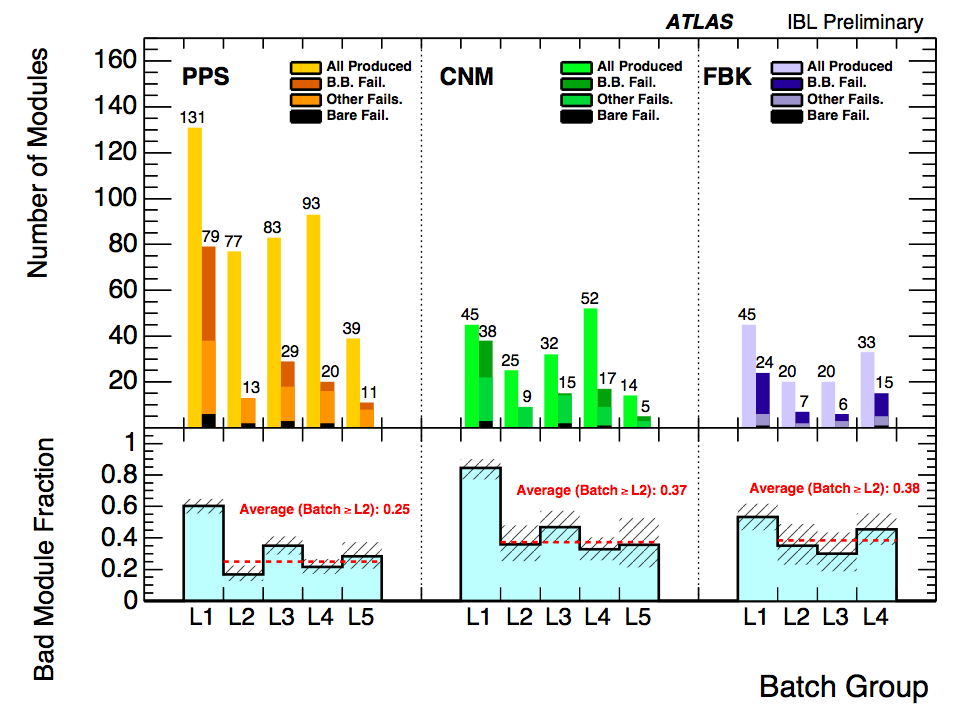}
\caption{Yield of IBL module production for module types of two-chip( Planar, as PPS) and single-chip (3D as CNM, and FBK) per production batch group. In the top panel for failure modules, "B.B. Fail." stands for large bump-bonding failure, "Bare Fail." stands for the module not assembled due to mainly mechanical damages, and "Other Fails." stands for both electrical and sensor failures discovered after assembly \cite{5}.}
\label{fig:IBL_module-prod}
\end{figure}

 A total of 710 modules have been produced in between January and October 2013. The thinning and the bump-bonding of the modules were carried out at IZM (Germany). The front-end chips were thinned to a thickness of 150\,$\mu$m and flip-chipped to the sensors using Sn\,Ag solder bumps. Then a double copper layer flex circuitry (so called module-flex) was glued onto the sensors and a wire bonding connection was done to electrically connect front-end chips and module flexes. Figure\,\ref{fig:IBL-modules} shows a photo of a dressed IBL two-chip (a) and a single-chip (b) module.  
All the produced modules have been mechanically, electrically and functionally qualified. The quality assurance procedure included short electrical and functional tests (e.g. IV-curves, threshold and noise scans) after the assembly at 15\,$\degree$C, thermal stress (ten cycles from -\,40\,$\degree$C to +\,40\,$\degree$C), radioactive sources tests and intensive electrical and functional tests and calibration at -\,15\,$\degree$C. During the modules production the first batch of both single- and two-chip modules had a large number of bump bonding failures which was traced back to the excessive flux in flip-chip process. Figure\,\ref{fig:IBL_module-prod} summarizes the yield of the produced modules for both sensor technologies, where a yield of  75 \% for the two-chip module, and  63\% for single-chip (CNM) and 62\% (FBK) modules. This production yields do not include the first production batch.
\par
\begin{figure}[h!]
\centering
\includegraphics[scale=0.32]{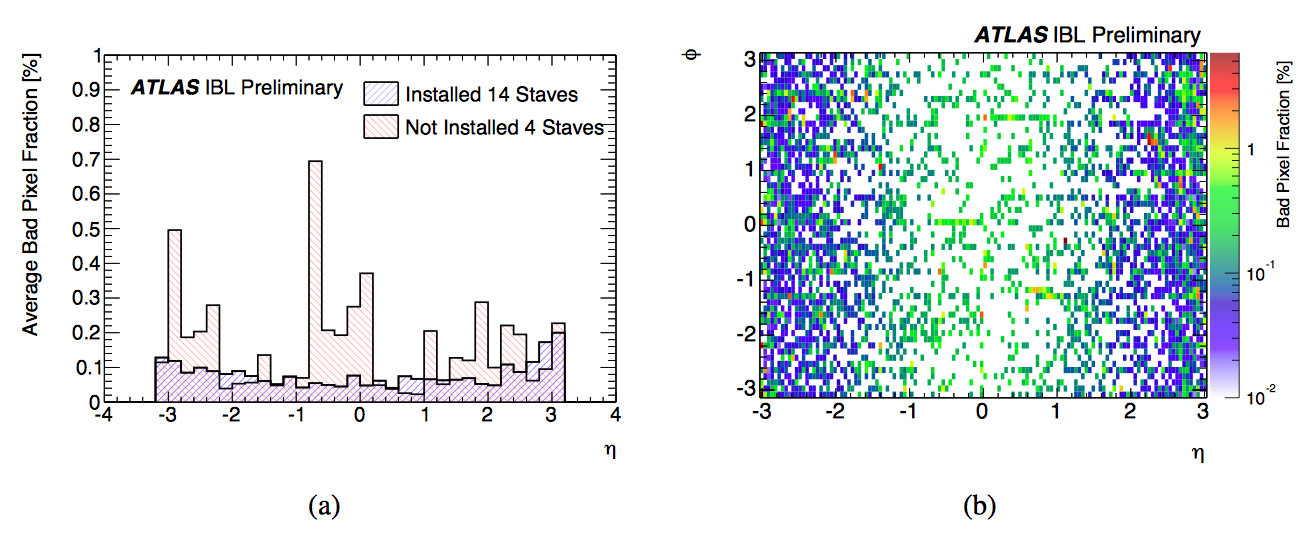}
\caption{(a) Average bad pixel ratio distribution as a function of $\eta$ for installed and not installed production staves and (b) in the $\eta$\,-\,$\phi$ plane for the 14 IBL staves \cite{7}.}
\label{fig:IBL_bad-pixels}
\end{figure}

\begin{figure}[h!]
\centering
\includegraphics[scale=0.55]{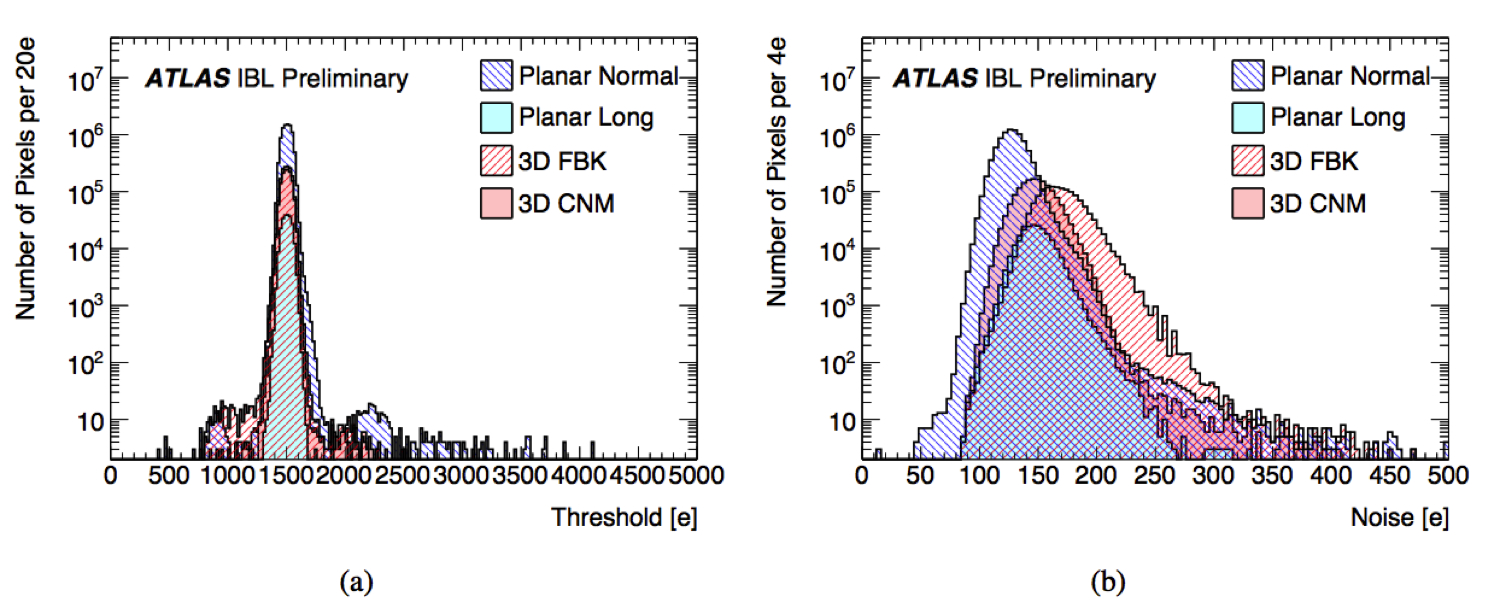}
\caption{(a) Threshold and (b) noise distributions after tuning all pixels to a target threshold of 1500\,e$^{-}$ at -\,12\,$\degree$C module temperature for the 14 IBL staves. The pixels are displayed in four categories according to normal and long pixels on planar sensors as well as pixels on FBK and CNM 3D sensors. \cite{7}.}
\label{fig:IBL_Th_Noise}
\end{figure}

 18 staves have been produced, qualified, and out of them the best 14 have been selected to be assembled in the IBL Detector. The quality assurance (QA) procedure allowed a detailed characterization of  all the staves including calibration in cold conditions and data taking with radioactive sources. The pixel defects were classified by type of failures as pertaining to the front-end chip, the sensor or the bump bonding. Figure\,\ref{fig:IBL_bad-pixels} shows the distribution of the bad pixel fraction for the produced staves, while Figure\,\ref{fig:IBL_Th_Noise} the threshold and  noise distributions after tuning all pixels to a target threshold of 1500\,e$^{-}$ at -\,12\,$\degree$C module temperature for the 14 IBL staves. 
\par In May 2014 the IBL Detector has been successfully installed inside ATLAS and then the commissioning started.  The first set of tests confirmed that the detector operated very stable at room temperature as well as at -12\,$\degree$C. Detailed comparisons between QA results and integration test results confirmed that the module operation are identical before and after the integration. In September 2014 the IBL Detector was partially included for the first time in the combined ATLAS cosmic runs (M5) reaching successfully  the 120\,kHz of readout capability. The next major steps are the commissioning of the combined Pixel and  IBL Detector system, tuning of  the detector and integration into the ATLAS data taking system.

\section{Summary}
The ATLAS Pixel Detector has shown excellent performances during the whole Run-1 of LHC. Then, taking advantage of the LHC long shutdown, ATLAS has used such period to consolidate the Pixel Detector with the replacement of the Pixel services, the repairing of the majority of failed modules and the installation of the new innermost pixel layer at an average radius of 33mm. With the upgraded 4-Layer Pixel Detector installed in the experiment, ATLAS will improve the vertex resolution, secondary vertex finding and b-tagging extending the reach of the physics analysis.

\end{document}